\documentclass[conference]{IEEEtran}
\IEEEoverridecommandlockouts
\usepackage{cite}
\usepackage{amsmath,amssymb,amsfonts}
\usepackage{algorithmic}
\usepackage{graphicx}
\usepackage{textcomp}
\usepackage{xcolor}
\def\BibTeX{{\rm B\kern-.05em{\sc i\kern-.025em b}\kern-.08em
    T\kern-.1667em\lower.7ex\hbox{E}\kern-.125emX}}
\usepackage{caption}
\usepackage{subcaption}

\begin{document}

\title{Enhancing 5G Performance: Reducing Service Time and Research Directions for 6G Standards\\
\thanks{This material is based upon a collaborative work between MIT and JMA Wireless. It is accepted for presentation at the 3rd edition of the International Conference on 6G Networking (6GNet 2024).}
}

\author{\IEEEauthorblockN{Laura Landon$^{\dagger}$, Vipindev Adat Vasudevan$^{\dagger}$, Jaeweon Kim$^{*}$, Junmo Sung$^{*}$, Jeffery Tony Masters$^{*}$, and Muriel M\'edard$^{\dagger}$}
\IEEEauthorblockA{
$^\dagger$Massachusetts Institute of Technology (MIT), Cambridge, USA, Emails: \{llandon9, vipindev, medard\}@mit.edu }
\IEEEauthorblockA{
$^*$JMA Wireless, Syracuse, USA, Emails: \{jkim, jsung, tmasters\}@jmawireless.com }
}

\maketitle

\begin{abstract}
    This paper presents several methods for minimizing packet service time in networks using 5G and beyond. We propose leveraging network coding alongside Hybrid Automatic Repeat reQuest (HARQ) to reduce service time as well as optimizing Modulation and Coding Scheme (MCS) selection based on the service time. Our network coding approach includes a method to increase the number of packets in flight, adhering to the current standard of the 16 HARQ process limit, demonstrating that these strategies can enhance throughput and reduce latency. Experimental results show that network coding reduces service times by up to $7\%$ in low SNR regimes, with greater reduction across all SNR as the number of packets in flight increases, suggesting that future 6G standards should consider increasing the number of HARQ processes for better performance.
\end{abstract}


\section{Introduction}
The development of 3GPP standards and research for each new generation is always heavily informed by lessons learned and innovations developed during the rollout of the previous generation. The 5G era introduced several innovations over LTE regarding reliability management. 
The turbo codes of LTE were largely replaced with Low-Density Parity-Check (LDPC) and polar codes. 
The hybrid ARQ systems of both LTE and 5G enhance reliability and throughput by combining error correction coding with retransmissions, but 5G adds a number of innovations to improve flexibility in scheduling, such as asynchronous downlink as well as uplink, and more dynamic scheduling of HARQ processes at the MAC layer \cite{3gpp.38.214, 3gpp.38.321}.

A key element of ensuring reliability and quality of service in 5G is the appropriate selection of modulation and coding scheme (MCS), which determines how many bits can be sent per symbol and how much redundancy is introduced to compensate for errors. Currently, MCS selection in 5G NR systems predominantly uses channel quality metrics such as signal-to-noise ratio (SNR) and predefined lookup tables to choose an MCS which will optimize data throughput \cite{lopez2018performance}. However, throughput is not always the primary consideration of a system, such as in the case of ultra-reliable low latency communications (URLLC), and minimizing the throughput does not necessarily minimize the delay.

Network coding is an approach, widely discussed outside the current 3GPP standard, for ensuring reliability of packets in a network within the stringent latency requirements of 5G use cases \cite{dias2023sliding}. In network coding, multiple packets are combined algebraically in groups of $K$ before being sent and additional $N-K$ packet combinations are added as forward erasure correction (FEC), with the result that the receiver can decode the necessary information with any combination of $K$ packets \cite{ahlswede2000network,ho2006random,fragouli2006network}. If the number of FEC redundant packets is sufficient to compensate for the lost packets, retransmissions are unnecessary. Thus network coding can provide a lower packet service time on average than HARQ where retransmissions are performed based on feedback after a round trip time (RTT). 

As standardization bodies actively discuss the standards and innovations for 6G, significant improvements are needed to meet the URLLC requirements of future applications. Through a comprehensive analysis of current HARQ-based reliability mechanisms, we propose several approaches incorporating network coding that achieve high reliability with low latency for next-generation reliability mechanisms. By building on the current 3GPP standards \cite{3gpp.38.214,3gpp.38.321} and addressing their challenges, we ensure compatibility with 5G while paving the way for the evolution to 6G systems.

Particularly, this work discusses multiple innovations to reduce service time and a method for choosing the MCS to minimize service time. We demonstrate that replacing HARQ with network coding reduces the service time. We also showcase that the relative benefit of network coding over HARQ with respect to packet service time increases with the number of packets that can be transmitted within a single round trip time, and propose novel methods of using network coding to increase the number of packets in flight at a given time within the bounds of the 16 HARQ process restriction in the current 3GPP standards. This method reduces the wait time for all SNR values while simultaneously maintaining a similar throughput.

\section{Related Work}
Low latency is a key concern in 5G networks, as evidenced by the introduction of URLLC, with its $99.999\%$ reliability and $1$ ms latency requirements \cite{3gpp.22.862}. Reliability and latency are often competing requirements, since improving reliability often introduces redundancies which increase the time required to complete transmission. Transmission time and its relation to the current reliability methods in the 3GPP standard, ARQ and HARQ, has been discussed in \cite{el2006mimo, ahmadi2010mobile, chelli2018throughput}. Performance of ARQ and HARQ generally improve with incorporation of techniques which adapt to the channel conditions \cite{towhidlou2018improved, ding2023harq}. One of these techniques is adaptive modulation and coding. 

Adaptive modulation and coding (AMC) is a technique used to dynamically adjust the MCS based on the current conditions of the wireless communication channel. Generally, the optimum MCS is determined according to previously defined SNR thresholds. These thresholds can be optimized for a variety of metrics. Throughput is commonly used, either with or without a maximum block error rate (BLER) constraint \cite{lopez2018performance}. Some studies have looked into using AMC to reduce delay, such as \cite{adhicandra2009optimizing} which found that more aggressive AMC tables reduce delay with HARQ in WiMAX. Others generated new tables for AMC by minimizing delay metrics defined in a variety of ways, such as slots needed to transmit normalized to the highest MCS in ideal conditions \cite{lopez2018performance}, overlap of consecutive transmissions in GPRS \cite{lopez2010link}, or time needed to transmit a collection of blocks in EGPRS \cite{luo2000packet}. 

Network coding is a technique for ensuring reliability which has been shown to have significant benefit in delay compared with other reliability techniques \cite{skulic2013delay, ao2012rate, lucani2012coding}. The techniques for AMC described above explore the selection of MCS for ARQ or HARQ, generally in earlier generations of the 3GPP standard. This paper adds to this area by exploring a way to minimize service time of transport blocks in 5G, and compares the results for HARQ vs network coding.

\section{Methodology} \label{sec:methodology}
\begin{figure}[!h]
     \centering
     \vspace{-0.5cm}
     \includegraphics[width=0.95\columnwidth]{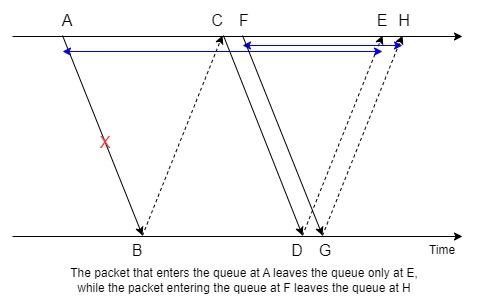}
     \caption{Service times for two packets.}
     \label{fig:wait_time2}
\end{figure}

In order to examine the improvement in packet service times, it is necessary to establish how we will calculate these service times. Service time corresponds to the time when a packet is first sent and a positive acknowledgment is received for the same (see Fig. \ref{fig:wait_time2}). For different communication techniques, such as ARQ, HARQ, and network coding, this varies depending on how lost packets are compensated. In selective repeat ARQ (SR-ARQ), the lost packets are retransmitted when a negative acknowledgment is received by the sender, until the packet is successfully received \cite{cipriano2010overview}. HARQ follows the same approach but with additional redundancies such that a packet can be recreated intact by combining partially corrupted packets using the LDPC codes. Both these techniques require at least an additional RTT to cover for the lost packet. Network coding uses forward erasure correction technique by sending repair packets a priori to compensate for the lost packets. In the case of network coding, any new reception provides a new degree of freedom, equivalent to a new packet in the ARQ scenario. This reduces the expected wait time. Further details of the service times calculations are explained below. Throughput is defined as the number of original packets sent per second. It can also be estimated as the reciprocal of average service time in a steady state. In case of a completed data transmission, it can also be computed as the total number of packets sent per the time taken to complete the transmission, a method used in most practical simulations.

\subsubsection{ARQ and HARQ} \label{HARQ}
The expected service time of a queue using a straightforward ARQ-only system would be given by 

\begin{equation}
\begin{aligned}
E[X] &= (1-p) \cdot RTT \\
&+ p (1-p) \cdot 2\times RTT \\
&+ p^2(1-p) \cdot 3\times RTT ...
\end{aligned}
\end{equation}

where $X$ is the service time of a packet, $p$ is the probability of erasure of that packet, and $RTT$ is the round trip time between sender and receiver. 

Note that we estimate the time from transmission to reception and error checking of a packet as $RTT$. At a bitrate of 1 Gbps or greater and an RTT of 10 ms, the propagation time of a transport block at maximum size (approximately 1 million bits) is on the order of a millisecond or less, an order of magnitude smaller than the RTT. The propagation time of a transport block only begins to approach the RTT when both the bitrate is lower than 1 Gbps and the transport block is at its maximum size. Since we are interested in the higher bitrates targeted by 6G \cite{chen2020vision}, we use the assumption that the propagation time can be neglected relative to the RTT to simplify our equations.  

5G NR systems use a combination of hybrid ARQ (HARQ) and ARQ to ensure reliable delivery of packets. The estimation of service time for 5G HARQ is complex because it uses a strategy of packet recombining to improve the likelihood that each subsequent transmission will result in a correctly received packet. This means that the probability of successful packet arrival increases with each retransmission, rather than remaining constant. 

The probability of erasure in an ARQ-only system, $p$, is estimated by the block error rate (BLER) which can be found experimentally or through simulations. This block error rate varies with SNR and MCS, meaning that $p$ can be expressed as a function of SNR and MCS, $p( MCS, SNR)$. 

Estimating the improved probability of each HARQ retransmission is complicated. In HARQ, the probability of successful reception on the first transmission is unchanged from that of an ARQ-only system, i.e. $p(SNR,MCS) = BLER(SNR, MCS)$. The probability of successful reception on the second transmission (first retransmission), however, is the sum of the probabilities of two different scenarios: first, that the second packet arrives perfectly intact, and second, that both the first and second packets arrive corrupted, but with sufficient information that they can be combined in HARQ to construct the correct packet. The first scenario is that of the ARQ-only scenario, and has probability $p*(1-p)$. However, in the second scenario, the probability of successfully recovery of a packet depends on the possibility of recovering the bits lost in each transmission through the different redundancy information, which can only be found by an empirical study. In this discussion, we use a simplified estimate of the expected value of the service time in a system using HARQ. The estimate is related to results obtained from using maximum ratio combining to see that a system with $N$ receive antennas and a given linear SNR value $s$ has the same effective signal to noise ratio as a system with only one receive antenna and a linear SNR of $N*s$. Now, in ARQ, each transmission of a packet is discarded if it is not received intact. In contrast, in HARQ, past transmissions of packets are saved and combined with new transmissions to reconstruct the correct packet. This process of combining transmissions received successively at a single antenna, although time-delayed, is effectively the same process as combining transmissions received simultaneously at multiple antennas.

To arrive at the derivation for our estimate of the probability of being able to decode a packet after N HARQ transmissions, we draw on concepts from estimation theory. It can be shown (see appendix) that for an AWGN channel, the variance of error on the estimate of a single transmission is 
\begin{equation}
    \sigma^2 = \frac{E[X^2]E[N^2]}{E[X^2] + E[N^2]}
\end{equation}  

and the variance of this error for N transmissions combined using HARQ is 
\begin{equation}
    \sigma^2 = \frac{E[X^2]E[N^2]}{N E[X^2] + E[N^2]}
\end{equation}

This is the same variance that would result from sending a single transmission with N times the power\footnote{This result is similar to the concept of maximum ratio combining (MRC).}.

We can use this result to guide our estimate of the probability of packet error at each successive retransmission using HARQ. If we assume that HARQ extracts the maximum mutual information from combining the packets received, then the probability of packet failure (which using our method depends on both MCS $\mu$ and SNR $s$) is $p(\mu,s)$ at the first transmission, $p(\mu,2s)$, at the second transmission, and $p(\mu, Ns)$ at the Nth transmission, where $s$ is in linear units before being multiplied by $N$. \footnote{One limitation of this method is its assumption that the channel conditions (affecting both MCS and SNR) remain the same across retransmissions. Refining our estimate to account for changing channel conditions or varying the MCS schemes will be considered in future works and here we assume channel conditions remain the same for brevity.}

Applying this to our equation of the expected service time, we get

\begin{equation} \label{eq:harq_ex}
\begin{split}
    E[X] &= (1-p(\mu , s)) \cdot RTT \\
    &+ p(\mu, s) (1-p(\mu , 2s)) \cdot 2\times RTT \\
    &+ p(\mu , s) p (\mu , 2s) (1-p(\mu , 3s)) \cdot 3\times RTT ...
\end{split}
\end{equation}


Note that this represents only one possible method for estimating the probability of packet failure at each retransmission using HARQ, and our proposed methods for minimizing service time in this document are not dependent on what method is used.

\subsubsection{Network Coding} \label{NC}

Network coding, as applied in this paper, is the use of erasure coding by nodes in the network to improve throughput and reliability. It is a technique wherein the information contained in packets is combined together before being transmitted, with additional redundant packet combinations sent at set intervals to compensate for any lost packets. The number of such repair packets can be defined based on the predicted probability of erasure in the communication channel. In contrast with ARQ or hybrid ARQ methods, network coding relies on this use of forward erasure correction to compensate for imperfections in the channel and seeks to avoid retransmission, although it is also possible to send additional repair packets based on feedback if required \cite{cohen2020adaptive}. Regardless of whether HARQ or network coding is used, similar levels of redundancy will be required to compensate for packet erasure, but by anticipating this redundancy with forward erasure correction, network coding reduces in-order packet delay by avoiding retransmissions and reduces the load on the network.

For those less familiar with the principles of network coding implementation, a node using the block network coding approach applied in this paper begins with a block of $K$ data packets to transmit. Rather than transmitting them directly, this node multiplies these $K$ packets by a $K \times N$ matrix of coefficients chosen from a finite field. This results in $N$ coded packets, each a linear combination of the original $K$, which are then transmitted by the node. The receiving node uses the inverse of the coefficient matrix to decode $K$ of the received packets. The key advantage here of network coding is that the receiver no longer needs to receive specific packets - any $K$ of the $N$ transmitted packets will suffice to decode the packets, because they are all a linear combination of the original packets. Thus the redundant $N-K$ packets can make up for the loss of any of the preceding $K$ packets. In order to optimize the level of redundancy, the value of $N$ is chosen such that $(N-K)/N$ approximates the block error rate. A more thorough discussion of network coding can be found in \cite{ho2008network,medard2012network}.

In a system using network coding, a packet loss is not necessarily a problem unless there are more losses in a block than there are redundant FEC packets. This means that in the worst case scenario, the system must wait until all packets in a block have been transmitted before it can determine whether retransmission is required. Now we are concerned not merely with the propagation time of a single packet, but of a whole block of packets, and this needs to be factored in while calculating the $E[X]$. 

We will represent the propagation time of a single packet as $\tau$. ($\tau$ can be estimated as $n_b / R_b$.) Then the expected service time of packets in a system using network coding with $K$ original packets and $N-K$ redundant packets (so $N$ total) can be calculated as follows:

\begin{equation} \label{eq:nc_ex}
    \begin{split}
        E[X] & \approx (1-p) \left (RTT + \tau \right ) \\
        &+  \sum_{i=1}^{N-K} {N \choose i} p^i (1-p)^{N-i} \left (RTT + i \tau + \frac{K+1}{2} \tau \right ) \\
        &+ \sum_{i=N-K+1}^{N} {N \choose i} p^i (1-p)^{N-i} \times \\ & \hspace{0.5cm}\left (RTT + \frac{2N - K + 1}{2} \tau +  RTT + (i-N+K) \times \tau \right )    
    \end{split}
\end{equation}

The first line of the equation represents the case that a packet arrives intact. The second line represents the case that a packet is lost, but there are sufficient redundant packets from network coding to make up for its loss. The last line represents the case that a packet is lost and that network coding is not sufficient to make up for it, so the lost packets must be retransmitted.

Note that equation \ref{eq:nc_ex} represents a case where the maximum number of retransmissions is 1. However, with network coding, additional redundancy can also be provided by increasing number of extra coded packets included in the block to avoid the retransmissions completely. This is particularly useful if the round trip time is significantly high.

\section{Methods to Minimize Service Time}\label{methods}

This paper explores several methods to minimize service time. These include a new metric for MCS optimization and two ways of network coding MAC layer transport blocks. 

\subsection{MCS Optimization}
The widely used approach of optimizing MCS code for throughput need not necessarily provide the lowest service time per packet. Service time per packet is an important metric to consider in mission-critical applications and in resource-constrained devices where the number of packets being kept in the service queue needs to be minimized. We propose that the average service time for a given round trip time can be used to optimize MCS for minimum service time. 

\subsection{Increasing Packets in Flight} \label{Hijack}
\begin{figure}[!h]
    \centering
    \includegraphics[width=0.95\columnwidth]{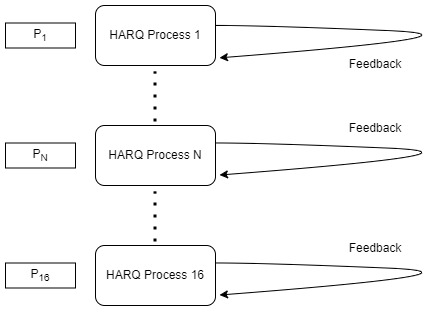}
    \caption{Limitation on number of packets due to HARQ process.}
    \label{fig:HARQ_process}
\end{figure}

As stated previously, the relative benefit of network coding over HARQ with respect to service times increases with the number of packets per round trip time. In 5G NR, the number of packets in flight is limited by the number of HARQ processes, capped at 16. Each HARQ entity will wait for feedback on the frame it sent, as shown in Fig.\ref{fig:HARQ_process}. 

\begin{figure}[!h]
    \centering
    \includegraphics[width=0.95\columnwidth]{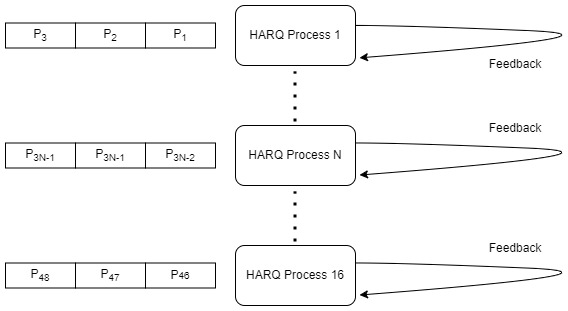}
    \caption{With network coding, the number of packets in flight can be increased by an order of how many original packets are included in a single code block.}
    \label{fig:NC_process}
     \vspace{-0.5cm}    
\end{figure}

However, network coding enables increasing the number of packets in flight, without disturbing the number of HARQ processes. With network coding, because any network-coded packet in a block can be used to decode any other and only the total number of received packets is necessary to decode an entire block,  feedback is only required for a block of packets rather than individual packets (in the form of the number of missing degrees of freedom). If each block of linear combinations of packets as described in section \ref{NC} is considered to be served using a single HARQ process, it allows each process to send multiple packets before getting any feedback, this eventually provides $K \times 16$ packets to be allowed in flight using the 16 HARQ processes, where $K$ is the size of original packets in a coded block. An example with a code where 3 original packets are included in each code block is shown in Fig. \ref{fig:NC_process}. Redundant packets are not explicitly shown in this figure because the key feature of this method is the use of block ACK/NACKs. Theoretically, it is also possible to implement network coding with no redundancy and it would still give the efficiency benefits of block acknowledgments. In this case the necessary number of repair packets can be sent based on the acknowledgment.

It is relevant to note that with this sending structure, the time of the coded block would be longer than the transmission time of a single packet, but this time will be negligible for the same reasons explained in the section \ref{sec:methodology}: propagation time vs RTT. This is because using the new model, a block of size $n$ will take $n \tau$ sec to transmit but only one RTT must be waited before receiving feedback and moving on to the next block, whereas in the existing design, each packet will take $\tau$ sec to transmit and an RTT before the process can move on, meaning that those same $n$ blocks would take $n(\tau + RTT)$ to process. Since we already established that RTT is significantly larger than tau, this means sending blocks is significantly more efficient than sending individual packets. The only concern here would be if $n$ is high enough that $n \tau$ begins to approach RTT, but even then, $n \tau + RTT << n (\tau + RTT)$ so sending blocks would still be more efficient than sending individual packets.

Another approach to increase the number of packets is by initiating multiple streams, by introducing an intermediate network coding layer between the RLC and MAC layers in the protocol stack. This network coding layer can take packets from the RLC layer, make them into multiple streams, and then code inside each stream. Each stream can have its own HARQ processes. This provides another layer of parallelization and further increases the number of packets in flight. This approach can work along with the block-level acknowledgment, providing a cascading effect in the number of packets in flight.

\begin{figure}[!h]
    \centering
    \includegraphics[width=0.95\columnwidth]{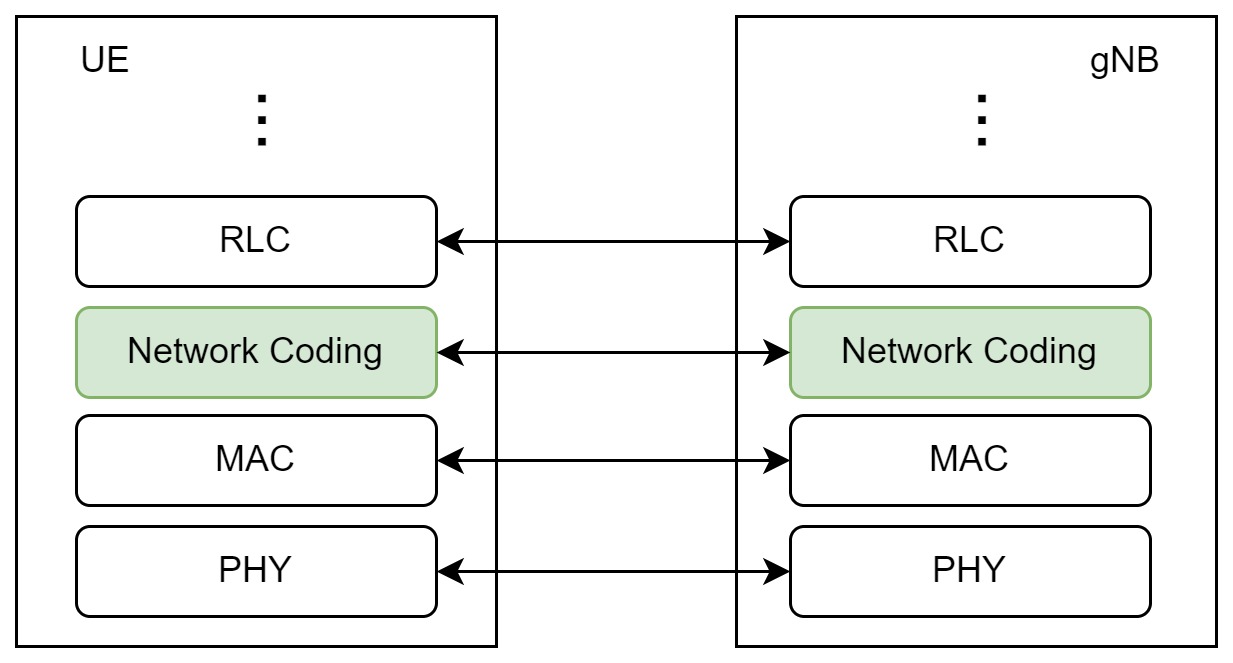}
    \caption{New network coding layer introduced in the 5G protocol stack.}
    \label{fig:NC_layer}
\end{figure}

As an aside on implementation methods, integrating network coding into a 5G system at the HARQ level requires MAC-level changes to the code on both the base station and the user equipment (UE), in order to both encode and decode packets. Commercially available base stations and UEs generally do not expose these lower levels of code for modification, but open source testbeds such as Eurecom's Open Air Interface are designed to be a publicly available implementation of 5G and can be used to access these lower layers. While we used Matlab to simulate the delay and throughput effects of network coding vs. HARQ, an implementation in such a 5G testbed would be the next logical step in establishing confidence in these results.

\section{Results and Discussion}
We collected simulated BLER values for each combination of MCS index and SNR value across a range of SNRs in a system with transport block retransmission disabled. We used MCS table 5.1.3.1-2 found in standard \cite{3gpp.38.214} and SNRs ranging from $-6$ to $27$ dB with increments of $0.1$ dB. Then using these BLER values as the probability of error $p$, we simulated the different reliability approaches (HARQ and network coding) in MATLAB to estimate the average service time for different approaches mentioned in section \ref{methods}. This simulation was performed on a slot by slot basis, where the transport block in any given slot was randomly determined to be in error or not according to a Bernoulli distribution with $p = BLER$. The service time is calculated as the time between the first transmission of a transport block and its eventual correct reception after retransmissions and/or coding depending on the reliability approach. Optimization of MCS based on service time was also performed. We further tested the performance of both approaches for the $99^{th}$ percentile of service times. This is particularly interesting as it helps provide service-level performance guarantees.

\subsection{Optimizing MCS for service time: HARQ}

\begin{figure}[!h]
    \centering
    \includegraphics[width=0.95\columnwidth]{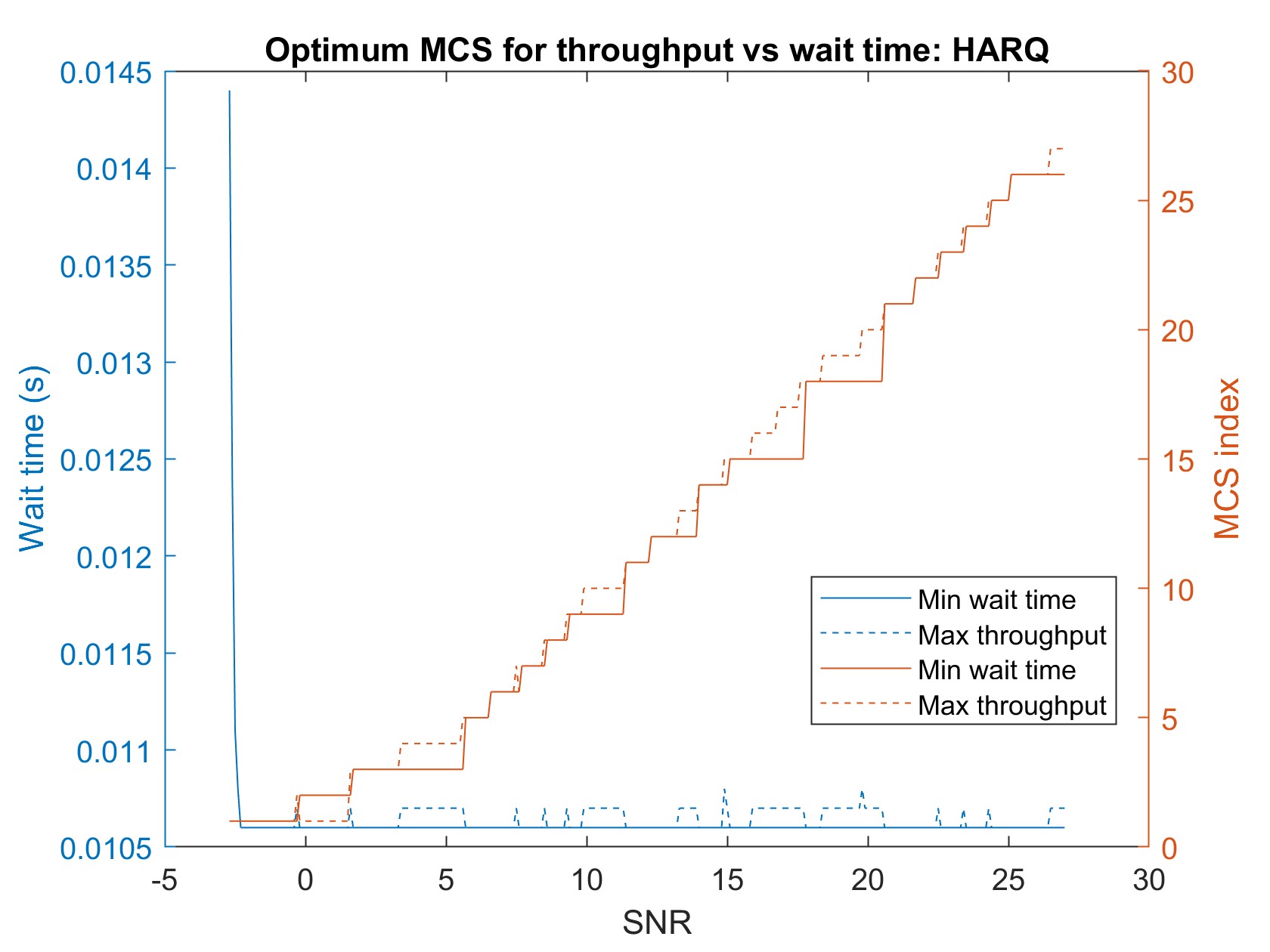}
    \caption{Optimizing MCS index for wait time vs. throughput for HARQ.}
    \label{fig:mcs_wt_v_tp}
\end{figure}

When there are multiple MCS options available for a particular $Eb/N0$, the optimal MCS for minimum service time can be different than the optimal MCS for maximum throughput. This can be seen in Fig. \ref{fig:mcs_wt_v_tp}, where the optimal MCS index for service time is different compared to the optimal MCS index for throughput. These different MCS indices result in different service times as well. With the current HARQ approach, the benefit of optimizing for service time is not providing significant performance gains. However, this may be important with a different coding scheme.

\subsection{Optimizing MCS for service time: Network coding reduces delay compared to HARQ}
When using service time-based optimization, it is particularly relevant to consider network coding. With sufficient and appropriate code rates, network coding provides a lower time in the queue than the current HARQ approach. The lower service time also corresponds to a higher number of packets being served, for a given arrival rate.

\begin{figure}[!h]
    \centering
    \includegraphics[width=0.95\columnwidth]{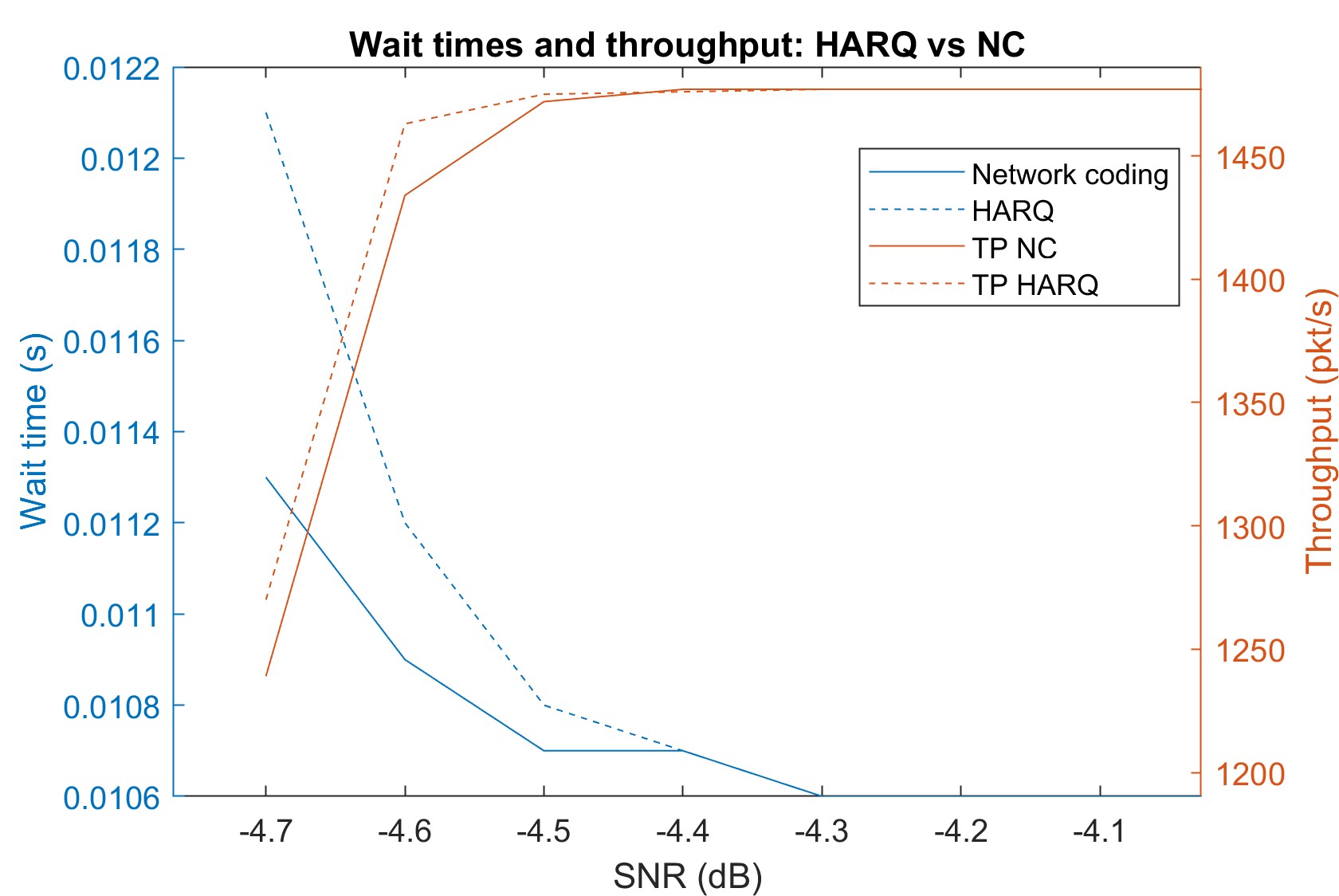}
    \caption{In the low SNR regime, network coding provides better wait times with only a small reduction in throughput.}
    \label{fig:low_SNR}
\end{figure}

The effects of network coding on service time optimization can be seen in Fig. \ref{fig:low_SNR}, in which it can be seen that network coding always results in a service time equal to or less than that of HARQ. With the limitation in 5G of 16 TBs maximum in flight over an RTT, this is particularly significant in low SNR regimes, where the erasure probability is higher.

\subsection{Limitation of 16 HARQ Processes}
The effect of transport blocks per round trip time on the disparity between service times for network coding vs. HARQ is shown below in Fig. \ref{fig:datarates}, in which the 16-process cap on HARQ is not taken into account in subfigure \ref{fig:160dr}.

\begin{figure}[ht]
    \centering
    \begin{subfigure}[b]{0.95\columnwidth}
        \centering
        \includegraphics[width=\columnwidth]{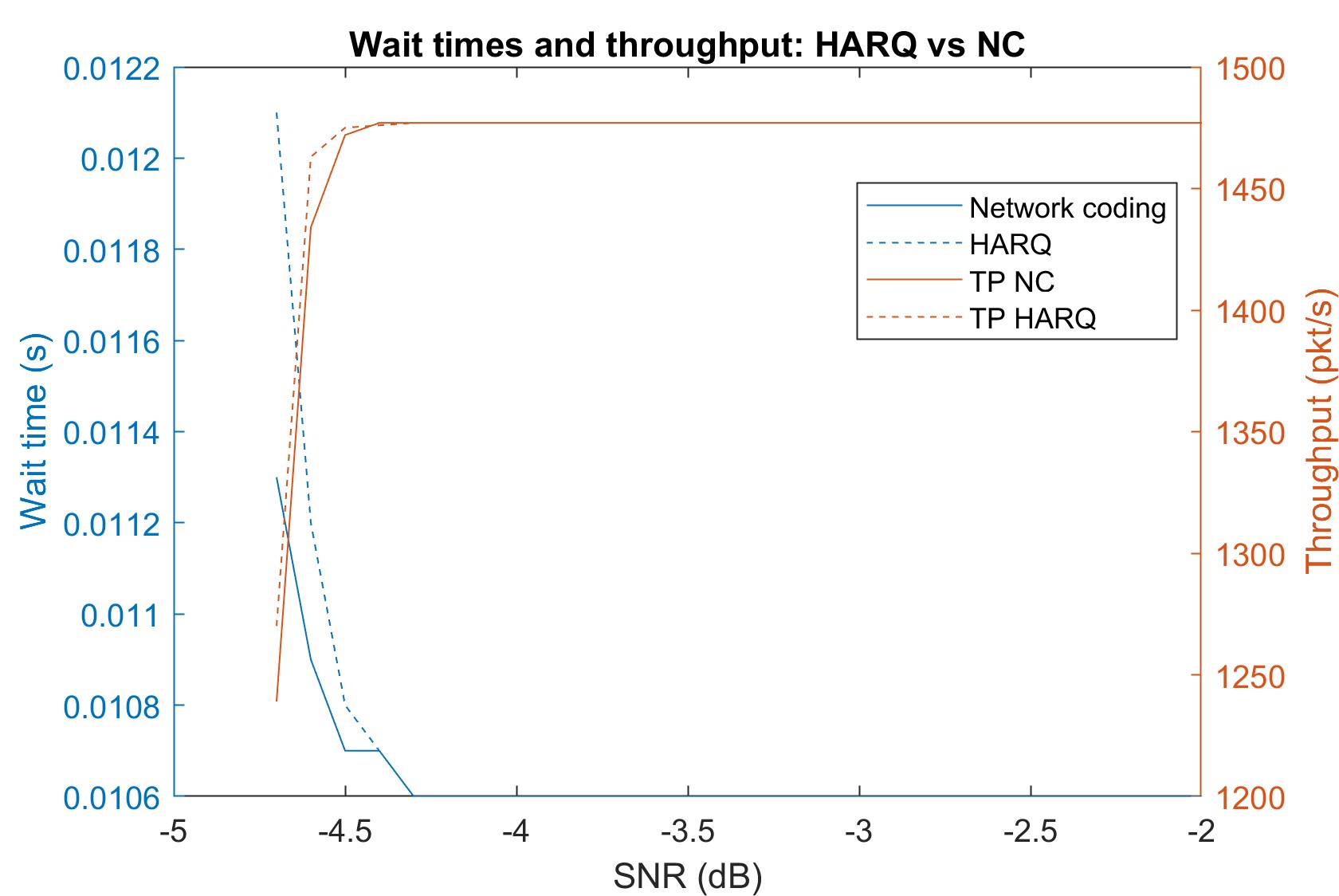}
        \caption{System in which 16 transport blocks are sent during one round-trip time.}
        \label{fig:16dr}
    \end{subfigure}
    \hfill
    \begin{subfigure}[b]{0.95\columnwidth}
        \centering
        \includegraphics[width=\columnwidth]{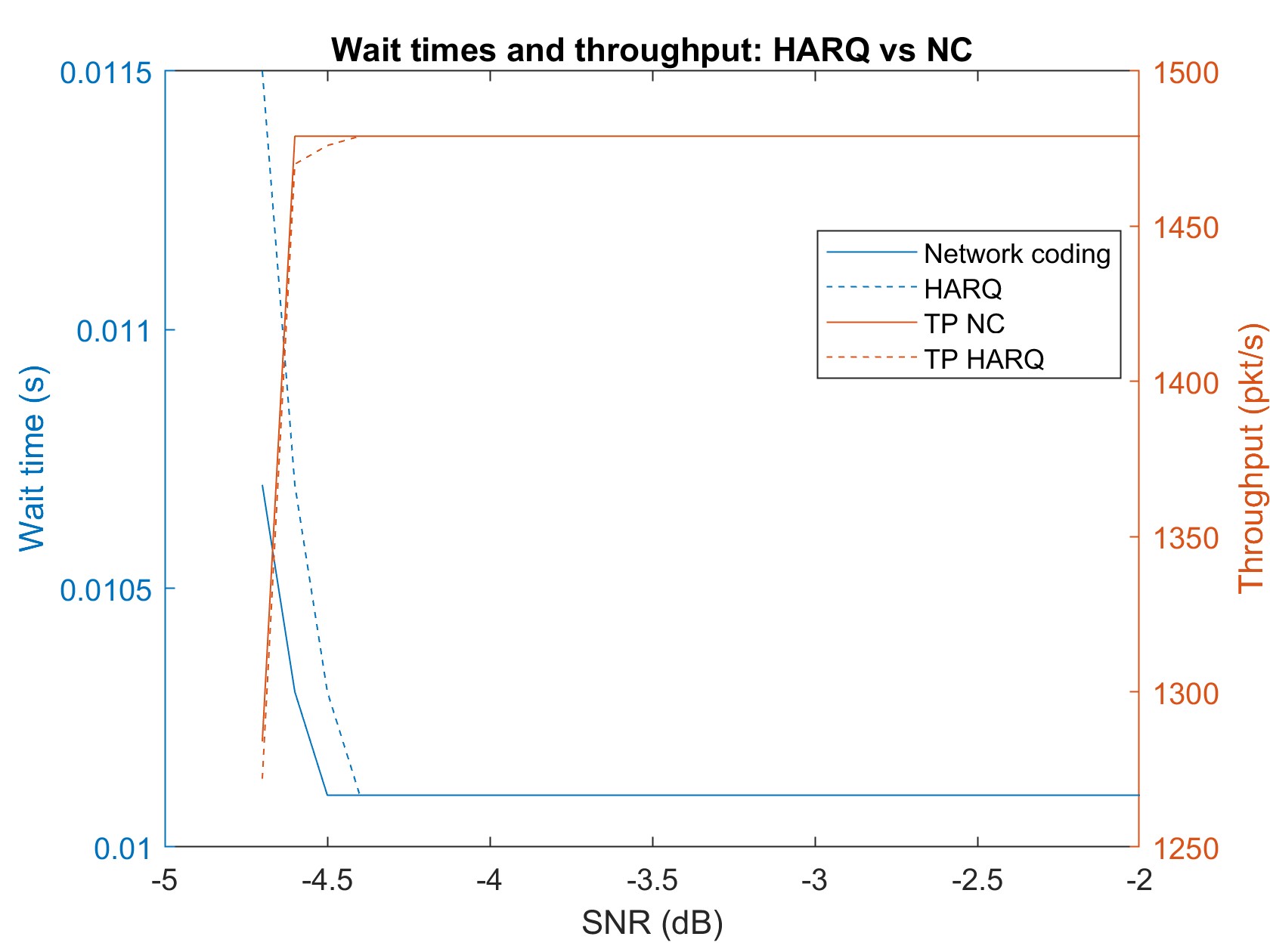}
        \caption{System in which 160 transport blocks are sent during one round-trip time.}
        \label{fig:160dr}
    \end{subfigure}
    \caption{Wait time comparison of systems with 16 vs. 160 transport blocks per RTT. Note that with 160 TBs per RTT, the wait time is reduced compared to 16 TBs per RTT.}
    \label{fig:datarates}
\end{figure}

These figures demonstrate that network coding has lower wait times than HARQ for both 16 and 160 packets per RTT, but the difference is more significant in the case of 160 packets per RTT. The current 16 packets per RTT limitation imposed by 5G standards means that this improvement cannot be directly applied to the current protocol stack, but future systems may find it worth exploring the benefit of increasing the number of packets that can be in flight during an RTT using the network coding layer.

\subsection{Increasing Packets in Flight via Network Coding}
\begin{figure}[!h]
    \centering
    \includegraphics[width=0.95\columnwidth]{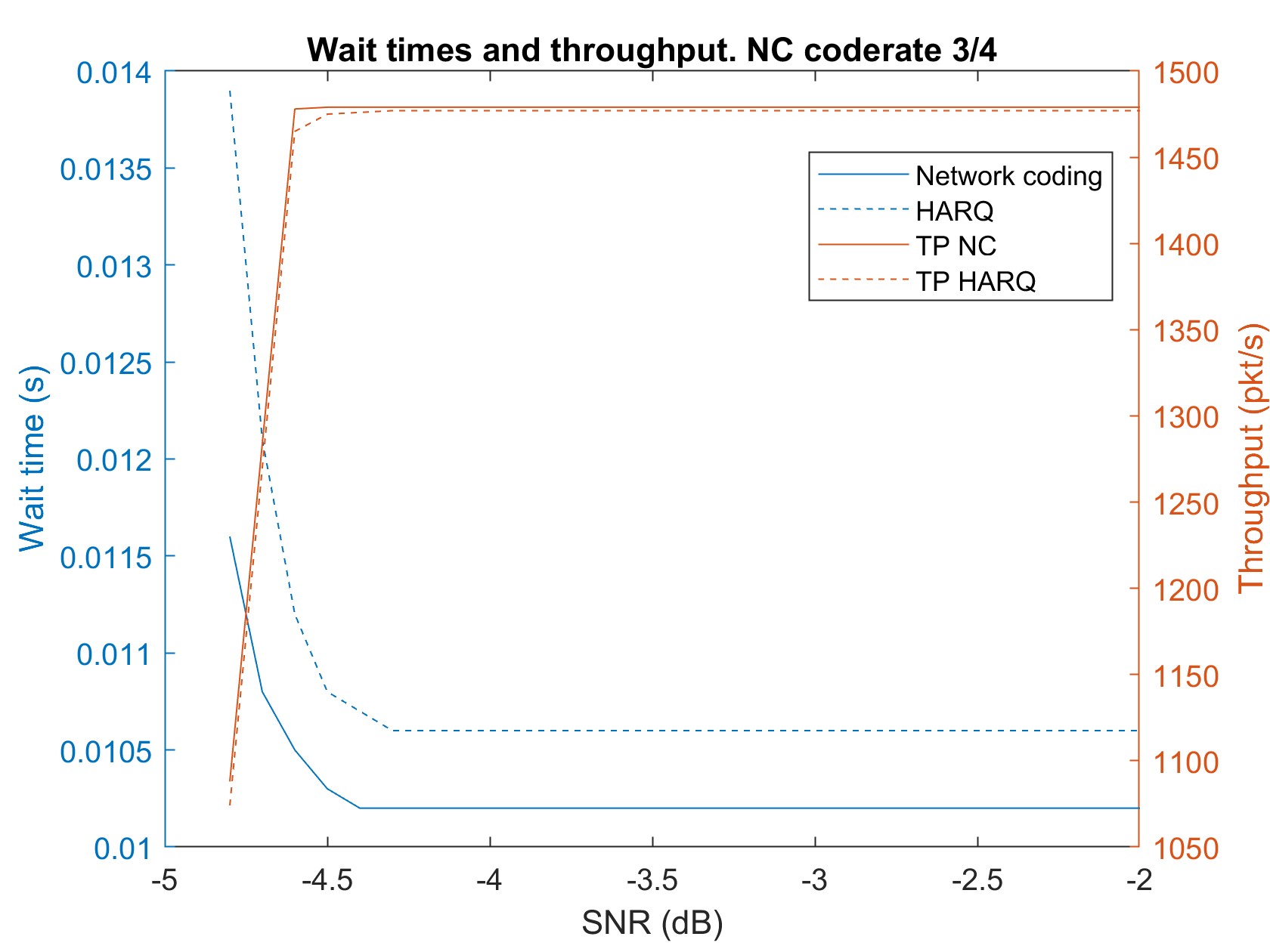}
    \caption{At a code rate of 3/4, network coding reduces wait times with no corresponding tradeoff in throughput.}
     \label{fig:hijack}
     \vspace{-0.5cm}
\end{figure}

While changes to the number of HARQ processes may be considered in future work, the method we proposed in subsection \ref{Hijack} to commandeer the HARQ process to send a complete network-coded block instead of a packet enables network coding to further improve the service time in a system while still working within the current 5G standard. With this approach, the improvement is consistent even in a high SNR range, where the probability of erasures is very low, as we can now send more packets. This also improves the throughput and provides better throughput when using network coding compared to HARQ. This approach is possible only with network coding, thanks to the possibility of working with block-level feedback. Fig.\ref{fig:hijack} shows the benefit of our approach with a code rate of $3/4$.

\subsection{Improved Service Level Agreement}
The improvement in service time provided by network coding has implications for service level agreements. For the same channel conditions, network coding has a lower $99^{\text{th}}$ percentile service time than HARQ. We define the $99^{\text{th}}$ percentile service time as the maximum of the lowest $99\%$ of service times for all transport blocks in a system. In other words, $99\%$ of transport blocks will have a service time less than or equal to the $99^{\text{th}}$ percentile service time. These $99^{\text{th}}$ percentile service times can be used as a service level guarantee.

\begin{figure}[!h]
    \centering
    \includegraphics[width=0.95\columnwidth]{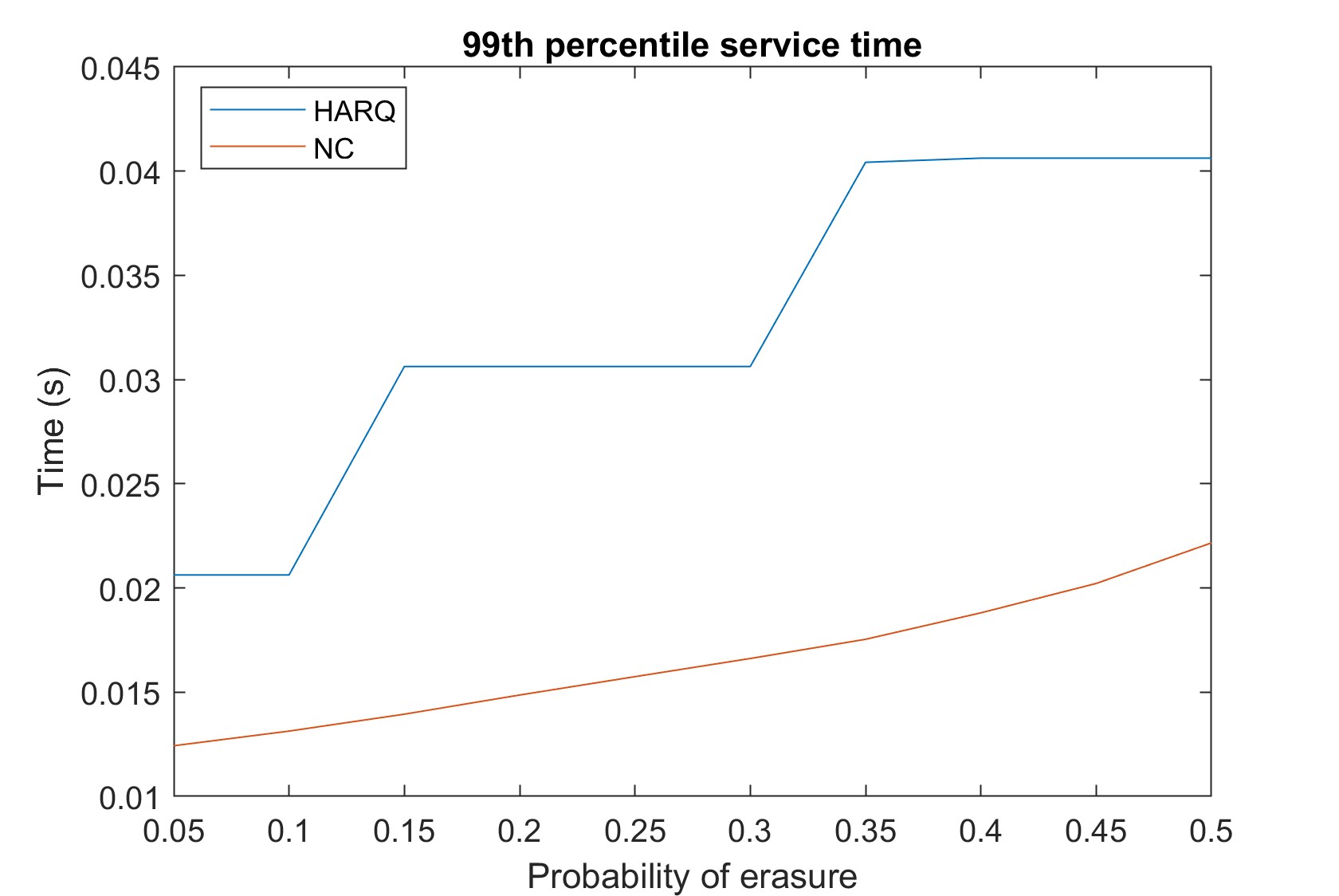}
    \caption{The $99^{th}$ percentile service times of HARQ are approximately double to those of NC.}
    \label{fig:SLA}
     \vspace{-0.5cm}
\end{figure}

Fig. \ref{fig:SLA} compares the service level guarantees of network coding and HARQ for different channel conditions, represented by the probability of erasure.

\section{Conclusion}
In this paper, we have investigated a method for optimizing Modulation and Coding Scheme (MCS) selection in 5G networks to minimize packet service time. Our analysis demonstrated that the packet service time can be reduced up to $7\%$ in low SNR regimes by integrating network coding with or in place of traditional HARQ mechanisms. Additionally, we proposed an innovative approach using network coding to increase the number of packets in flight within the current constraint of 16 HARQ processes, demonstrating that packet service times can be reduced for all SNRs without reducing the throughput. Future work could explore applying network coding to TDD. As network coding requires a feedback only per block, not per frame, the number of feedback slots in a TDD pattern can be reduced. This allows more packets to be sent within a time frame, another advantage of network coding that can result in increased throughput. While our methods show considerable improvements within the existing 5G framework, our findings also suggest that future wireless standards, such as 6G, could benefit from increasing the number of HARQ processes beyond the current limit. This enhancement would enable more flexible and efficient data transmission strategies, further reducing packet service times and improving overall network performance.

\appendix

To arrive at the derivation for our estimate of the probability of being able to decode a packet after N HARQ transmissions, we consider tools from estimation theory. Rather than asking whether a packet was intact or corrupted, we will treat any received packet as a measured value $Y$ of the sent packet, $X$. We have $Y = X + N$ where $N$ represents the effects of noise.

We will represent the channel as AWGN. This means that the minimum mean-squared error (MMSE) is also the linear least square error (LLSE). Then we are seeking to minimize the squared error, 
\begin{equation}
\begin{aligned}
  E[(\hat{X} - X)^2] &= E[(\alpha Y_1 - X)^2] \\
  &= E[(\alpha X + \alpha N - X)^2] \\
  &= E[(\alpha - 1)^2 X^2 + 2 \alpha (\alpha - 1) X N + \alpha^2 N^2] 
\end{aligned}
\end{equation}
The middle term $2 \alpha (\alpha - 1)XN$ has an expectation of zero because the signal and noise are independent and zero mean. Since the expectation term is a linear operator, this gives
\begin{equation}
\begin{aligned}
  E[(\hat{X} - X)^2] &=\\ (\alpha - 1)^2 E[X^2] + \alpha^2 E[N^2].
\end{aligned}
\end{equation}
We can take the derivative of the $\alpha$ term from here to find that this error expectation is minimized at
\begin{equation}
  \alpha = \frac{E[X^2]}{E[X^2] + E[N^2]}.
\end{equation}

Now, consider the case of a single HARQ retransmission, so that we have received the same packet transmitted twice. In this case, we would like to combine the information we have received from these two measurements and form one single estimate. This can be represented as $\hat{X} = \alpha_1 Y_1 + \alpha_2 Y_2$, where $Y_1$ and $Y_2$ are the two received transmissions and $\alpha_i$ gives the corresponding weighting. Note that $Y_i = X + N_i$, with $X$ the same in each case because it is the same packet being transmitted. In this case, we can repeat the process shown above to see that 
\begin{equation}
    \begin{split}
    E[(\hat{X} - X)^2] &= E[(\alpha_1 Y_1 + \alpha_2 Y_2 - X)^2] \\
    &= E[(\alpha_1 X + \alpha_1 N_1 + \alpha_2 X + \alpha_2 N_2  - X)^2] \\
    &= (\alpha_1 + \alpha_2 - 1)^2 E[X^2] + \\ & \hspace{1cm} \alpha_1^2 E[N_1^2] + \alpha_2^2 E[N_2^2].
    \end{split}
\end{equation}
  
We can take two partial derivatives (one with respect to $\alpha_1$ and one with respect to $\alpha_2$) to find that this error expectation is minimized at 
\begin{equation}
    \alpha_1 = \frac{(1 - \alpha_2)E[X^2]}{E[X^2] + E[N^2]}
\end{equation}
and 
\begin{equation}
  \alpha_2 = \frac{(1 - \alpha_1)E[X^2]}{E[X^2] + E[N^2]}.
\end{equation}

Solving for $\alpha_1$ and $\alpha_2$ will show that $\alpha_1 = \alpha_2$, so we drop the subscripts and obtain
\begin{equation}
  \alpha = \frac{\frac{E[X^2]}{E[X^2] + E[N^2]}}{1+\frac{E[X^2]}{E[X^2] + E[N^2]}} = \frac{E[X^2]}{2E[X^2] + E[N^2]}.
\end{equation}

Now, the variance in the case of a single transmission is 
\begin{equation}
  \sigma_1^2 = \frac{E[X^2]E[N^2]}{E[X^2] + E[N^2]}
\end{equation}
and the variance in the case of two transmissions is 
\begin{equation}
  \sigma_2^2 = \frac{E[X^2]E[N^2]}{2E[X^2] + E[N^2]}.
\end{equation}
It can similarly be shown that this generalizes for $K$ transmissions, such that 
\begin{equation}
  \sigma_K^2 = \frac{E[X^2]E[N^2]}{KE[X^2] + E[N^2]},
\end{equation}
which is the same variance which is obtained through maximal ratio combining \cite{medard1999processing}, a method for diversity combining which adds the signals from each channel together.

Now, the mutual information is related to the entropy by the equation $I(X; (Y_1, ... , Y_K)) = H(X) - H(X | Y_1, ... , Y_K)$. Note that variance can be lower bounded as follows:
\begin{equation}
  \sigma_K^2 = E[(\hat{X} - X)^2] \geq \frac{1}{2 \pi e} 2^{2h(X | Y_1, ... , Y_K)} 
\end{equation}
and equality holds if $X$ is Gaussian \cite{fang2019generic}. In this case we have $h$ minimized as $h(X | Y_1, ... , Y_K) = \frac{1}{2} + \frac{1}{2} \ln \left (2 \pi \sigma_K^2 \right )$,
which means that $I(X; (Y_1, ... , Y_K))$ is maximized when we set $X$ as Gaussian. Then the maximum mutual information of the transmissions depends solely on the variance in the error estimate. Since the variance we obtain by combining the information content of successive transmissions is the same as the variance obtained using maximal ratio combining \cite{medard1999processing}, we see that the maximal mutual information of the tranmissions must also be the same as that obtained using maximal ratio combining.

\bibliographystyle{IEEEtran}
\bibliography{IEEEabrv,references}

\end{document}